\documentclass[letterpaper]{article} 
\usepackage{aaai2026}  
\usepackage{times}  
\usepackage{helvet}  
\usepackage{courier}  
\usepackage[hyphens]{url}  
\usepackage{graphicx} 
\usepackage{natbib}  
\usepackage{caption} 
\usepackage{algorithm}
\usepackage{algorithmic}
\usepackage{threeparttable}
\usepackage{booktabs}
\usepackage{multirow}
\usepackage{amssymb} 
\usepackage{mathrsfs}
\usepackage{amsmath}
\usepackage{amsmath}
\usepackage{xcolor}
\newtheorem{theorem}{Theorem} 
\usepackage{newfloat}
\usepackage{listings}
\DeclareCaptionStyle{ruled}{labelfont=normalfont,labelsep=colon,strut=off} 
\floatstyle{ruled}
\newfloat{listing}{tb}{lst}{}
\floatname{listing}{Listing}
\title{Transferable Hypergraph Attack via Injecting Nodes into Pivotal Hyperedges}
\author {
    Meixia He\textsuperscript{\rm 1},
    Peican Zhu\textsuperscript{\rm 1}\thanks{P. Zhu, K. Tang, Y. Guo are joint corresponding authors.},
    Le Cheng\textsuperscript{\rm 1,2},
    Yangming Guo\textsuperscript{\rm 3}\footnotemark[1],
    Manman Yuan\textsuperscript{\rm 4},
    Keke Tang\textsuperscript{\rm 5,6}\footnotemark[1]
}
\affiliations {
    \textsuperscript{\rm 1}School of Artificial Intelligence, Optics and Electronics (iOPEN), Northwestern Polytechnical University\\
    \textsuperscript{\rm 2}School of Computer Science, Northwestern Polytechnical University\\
    \textsuperscript{\rm 3}School of Cybersecurity, Northwestern Polytechnical University\\
    \textsuperscript{\rm 4}School of Computer Science, Inner Mongolia University\\
    \textsuperscript{\rm 5}Cyberspace Institute of Advanced Technology, Guangzhou University\\
    \textsuperscript{\rm 6}Huangpu Research School of Guangzhou University\\   
    meixia\_he@mail.nwpu.edu.cn, ericcan@nwpu.edu.cn, chengle@mail.nwpu.edu.cn, \\ yangming\_g@nwpu.edu.cn, yuanman@imu.edu.cn, tangbohutbh@gmail.com
}

\usepackage{bibentry}

\begin{document}

\maketitle

\begin{abstract}
Recent studies have demonstrated that hypergraph neural networks (HGNNs) are susceptible to adversarial attacks. However, existing methods rely on the specific information mechanisms of target HGNNs, overlooking the common vulnerability caused by the significant differences in hyperedge pivotality along aggregation paths in most HGNNs, thereby limiting the transferability and effectiveness of attacks. In this paper, we present a novel framework, i.e., \textbf{T}ransferable \textbf{H}ypergraph \textbf{Attack} via Injecting Nodes into Pivotal Hyperedges (TH-Attack), to address these limitations. Specifically, we design a hyperedge recognizer via pivotality assessment to obtain pivotal hyperedges within the aggregation paths of HGNNs. Furthermore, we introduce a feature inverter based on pivotal hyperedges, which generates malicious nodes by maximizing the semantic divergence between the generated features and the pivotal hyperedges features. Lastly, by injecting these malicious nodes into the pivotal hyperedges, TH-Attack improves the transferability and effectiveness of attacks. Extensive experiments are conducted on six authentic datasets to validate the effectiveness of TH-Attack and the corresponding superiority to state-of-the-art methods.
\end{abstract}


\section{Introduction}
With real-world networks become increasingly complex and diverse, higher-order structures such as hypergraphs have emerged as powerful tools for encapsulating intricate interaction information within graph data \cite{battiston2021physics,antelmi2023survey,jin2019robust}. Furthermore, to effectively capture higher-order features within hypergraphs, Hypergraph Neural Networks (HGNNs) have been introduced \cite{feng2019hypergraph,bai2021hypergraph,kim2024survey,li2025deep}, outperforming Graph Neural Networks (GNNs) \cite{kipf2016semi,cheng2024gin,velivckovic2017graph,cheng2024heuristic,he2025plgnn,wang2025elevating} in capturing comprehensive feature information and achieving remarkable results across various downstream graph tasks like node classification, source detection etc. Despite their success, HGNNs have been shown to be as susceptible to adversarial attacks as GNNs \cite{hu2023hyperattack,chen2023momentum,hu2023hyperattack,chen2023momentum,wei2020adversarial}, raising concerns about their security.
\begin{figure}[t]
  \centering 
  \includegraphics[width=1\linewidth]{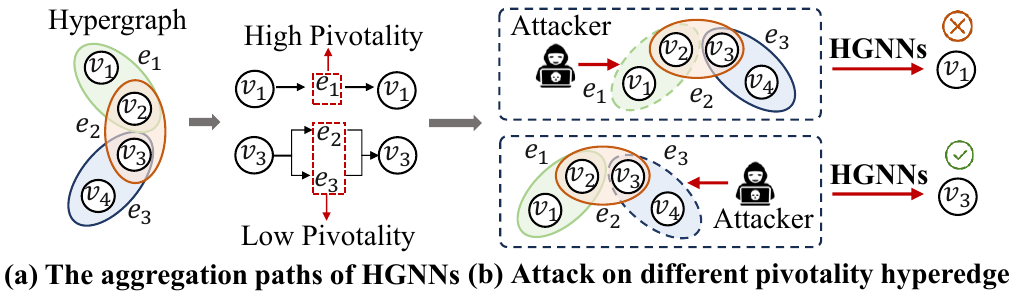}
  \caption{An illustration of motivation. (a) In HGNNs, the pivotality of hyperedges along the information aggregation paths varies significantly. (b) Attacking hyperedges with high pivotality significantly impacts HGNN performance, resulting in incorrect predictions for nodes that depend on this hyperedge for feature information.}
  \label{motivation}
\end{figure}

Attacks against GNNs could be divided into modification attacks and injection attacks based on the attack methodology. Modification attacks \cite{chang2020restricted,jin2023local,zhu2024simple} alter node/edge attributes or connections to degrade GNN performance, while injection attacks \cite{sun2019node,zou2021tdgia,zhang2024maximizing,zhu2024node} introduce malicious nodes/edges into the original graph. However, the higher-order relationships in hypergraphs prevent direct application of graph attacks to HGNNs. Consequently, hypergraph attacks against HGNNs are divided into hypergraph modification attacks and hypergraph injection attacks. For instance, HyperAttack \cite{hu2023hyperattack} modifies hyperedge connections using gradient guidance, and MGHGA \cite{chen2023momentum} uses surrogate models for untargeted feature modifications, both focusing on gradient-based modification attacks. Given that injection attacks demand lower privileges and are more practically implementable, recent strategies have emerged, including IE-Attack \cite{he2025hypergraph} and H3NI \cite{shi2025h3ni}. IE-Attack proposed injecting homogeneous nodes generated through Kernel Density Estimation (KDE) into elite hyperedges. H3NI \cite{shi2025h3ni} employed genetic algorithms to select the hyperedges that injected into the nodes. 

Although existing hypergraph attacks have achieved some advancements, they fail to find the common vulnerability caused by the significant differences in hyperedge pivotality along aggregation paths in most HGNNs, as illustrated in Figure \ref{motivation}(a). For instance, node $v_{1}$ aggregates higher-order features solely through hyperedge $e_{1}$, and node $v_{3}$ has two distinct aggregation paths. Therefore, attacking the $e_{1}$ would directly disrupt its information transmission in this path, rendering the HGNNs incapable of accurately predicting $v_{1}$, as illustrated in Figure \ref{motivation}(b). In contrast, node \(v_{3}\) has two aggregation paths, enabling it to retain partial true feature information even if \(e_{3}\) is attacked. Thus, attacking \(e_{3}\) does not affect HGNN and the correct result is still predicted. This illustrates that hyperedge $e_{1}$ exhibits a higher pivotality compared to hyperedges $e_{2}$ and $e_{3}$. Moreover, existing attack methods typically attack hyperedges with low pivotality, neglecting the common vulnerability where most HGNNs have pivotal hyperedges in their information aggregation paths. This limits their transferability and effectiveness to HGNNs with different architectures.

To tackle these challenges, we propose the \textbf{T}ransferable \textbf{H}ypergraph \textbf{Attack} via Injecting Nodes into Pivotal Hyperedges (TH-Attack). First, we analyze the pivotality of hyperedges in the information aggregation paths of HGNNs and design a hyperedge recognizer via pivotality assessment to identify pivotal hyperedges. Considering the propagation characteristics of pivotal hyperedges within information aggregation paths, we develop a feature inverter that generates malicious nodes by maximizing semantic divergence from pivotal hyperedges. Finally, by injecting malicious nodes into pivotal hyperedges, the structural integrity of these hyperedges is compromised. This disruption impedes feature propagation along information aggregation paths, resulting in information aggregation failure across most HGNNs and significantly degrading their performance. We validate the effectiveness of TH-Attack on six publicly available datasets. Extensive experimental results demonstrate that TH-Attack achieves superior attack performance and transferability across multiple HGNNs, outperforming state-of-the-art hypergraph injection attack methods.

Overall, our contributions are summarized as:
\begin{itemize}
    \item We are the first to identify and formalize the common vulnerability caused by significant differences in hyperedge pivotality along aggregation paths in most HGNNs.
    \item We propose the transferable hypergraph attack by injecting malicious nodes into the pivotal hyperedges located on information aggregation paths.
    \item We demonstrate the transferability and effectiveness of our proposed method over baseline approaches through extensive experimental validation.
\end{itemize}
\section{Related Work}
\subsubsection{Hypergraph Neural Networks}
To effectively model the higher-order interactions in complex systems, the hypergraph \(\mathcal{G}=(\mathcal{V},\mathcal{E})\) has been proposed and widely adopted in graph data analysis \cite{jin2019robust,antelmi2023survey}. Building upon this foundation, Hypergraph Neural Network (HGNN) \cite{feng2019hypergraph} was proposed to capture more comprehensive feature representations compared to Graph Neural Networks (GNNs). By incorporating techniques such as hypergraph convolution \cite{bai2021hypergraph}, hypergraph attention \cite{yadati2019hypergcn}, and adaptive hyperedge modeling \cite{zhang2019hyper}, the feature aggregation process in HGNNs had been significantly refined, demonstrating exceptional performance in tasks like node classification. Most of these methods had a common feature of satisfying the node-hyperedge-node feature aggregation mechanism, effectively overcoming the representation limitations of traditional graph structures. The feature extraction capabilities of HGNNs have driven their adoption across diverse domains, with large-scale deployments in applications such as recommendation systems \cite{wang2018dual,zeng2023multi} and biological networks \cite{yu2023basket}.

As HGNNs are increasingly used in critical areas like medical diagnosis and financial risk, examining their adversarial robustness has become an urgent security concern.

\subsubsection{Adversarial Attack against HGNNs}
While adversarial attacks on graphs had matured with methods ranging from modification to injection \cite{chang2020restricted,jin2023local,wang2020scalable,ju2023let,tao2021single,tao2023adversarial,wang2025stealthy}, attacking HGNNs demanded distinct strategies due to their intrinsic higher-order interactions. This has led to two analogous but structurally divergent attack categories: hypergraph modification attacks and injection attacks. For instance, the HyperAttack \cite{hu2023hyperattack} modified hyperedge link states of target nodes using gradient guidance, while MGHGA \cite{chen2023momentum} targeted untargeted feature modifications during the training phase via surrogate models. Both relied on gradient-based modifications of hyperedge structures. Given the lower privilege requirements and feasibility of injection attacks, methods like IE-Attack \cite{he2025hypergraph} and H3NI \cite{shi2025h3ni} were developed. IE-Attack generated homogeneous nodes via Kernel Density Estimation (KDE) for injection into elite hyperedges, enhancing stealth and effectiveness through hyperedge group identity. H3NI introduced a black-box node injection, using genetic algorithm and budget models to tackle hyperedge selection challenges specific to hypergraphs.

However, existing hypergraph attack methods did not explore varying pivotality of hyperedges in the information aggregation paths of HGNNs, limiting the transferability and effectiveness of attack methods.
\section{Preliminary and Problem Statement}
\subsubsection{Hypergraph Neural Networks}
A hypergraph, represented as \(\mathcal{G} = (\mathcal{V}, \mathcal{E})\), extends the concept of a standard graph \(G = (V, E)\) by permitting hyperedges \(e_{j} \in \mathcal{E}\) to connect two or multiple nodes, such as \(e_{2} = \{v_{2}, v_{3}\}\) shown in Figure \ref{frame}. The number of nodes is $N$, and the number of hyperedges is $M$. The node features are represented by the matrix $\mathcal{X} \in \mathbb{R}^{|\mathcal{V}|\times|\mathcal{F}|}$, where $|\mathcal{F}|$ is the feature dimension. The structure of the hypergraph is captured by its incidence matrix $H \in \mathbb{R}^{|\mathcal{V}| \times |\mathcal{E}|}$, defined as:
\begin{align}\label{H}
H_{ij} = \left\{
\begin{aligned}
  &1,~~~~ \text{if } v_{i} \in e_{j}, \\
  &0,~~~~ \text{if } v_{i} \notin e_{j}.
\end{aligned}
\right.
\end{align}

The feature aggregation of HGNNs at $l$-th layer is represented as follows:
\begin{align}\label{HGNN_eq}
\mathcal{X}^{(l+1)} = \sigma\Big( D_{\mathcal{V}}^{-\frac{1}{2}} H \mathcal{W} D_{\mathcal{E}}^{-1} H^{\mathsf{T}} D_{\mathcal{V}}^{-\frac{1}{2}} \mathcal{X}^{(l)} \Theta^{(l)} \Big),
\end{align}
where $D_{\mathcal{V}}$ is the node degree matrix with elements $D_{\mathcal{V}, ii} = \sum_{j} \mathcal{W}_{jj} H_{ij}$, $D_{\mathcal{E}}$ is the hyperedge degree matrix with elements $D_{\mathcal{E}, jj} = \sum_{i} H_{ij}$, $\mathcal{W}$ is the hyperedge weight matrix, $\Theta^{(l)}$ is the layer-specific trainable weight parameters for $l$-th layer, and $\sigma(\cdot)$ denotes a non-linear activation function.
\subsubsection{Hypergraph Node Injection Attack}  
Hypergraph Node Injection Attacks compromise HGNNs by injecting $m$ malicious nodes $\mathcal{V}_{mal} = \{v_{mal}^1, \dots, v_{mal}^m\}$ into existing hyperedges of the original hypergraph $\mathcal{G}=(\mathcal{V},\mathcal{E})$. This approach preserves legitimate nodes and hyperedges while strategically injecting nodes into existing hyperedges, augmenting them with carefully crafted features $\mathcal{X}_{mal}$. The attacked hypergraph becomes $\hat{\mathcal{G}} = (\hat{\mathcal{V}}, \hat{\mathcal{E}})$, where $\hat{\mathcal{V}} = \mathcal{V} \cup \mathcal{V}_{mal}$ and $\hat{\mathcal{E}} = { e_{j} \cup v_{mal}^{m} \mid e_{j} \in \mathcal{E} }$ denoting nodes injected into hyperedge, such as $e_{j}=\{v_{1},v_{2}\}\rightarrow\{v_{1},v_{2},v_{mal}^{m}\}$. Expanded incidence matrix is $\hat{H} \in \mathbb{R}^{(|\hat{\mathcal{V}}|+m) \times |\hat{\mathcal{E}}|}$, where $\hat{H}_{(|\hat{\mathcal{V}}|+m)j}=1$ indicate injected nodes $v_{mal}^{m} \in \mathcal{V}_{mal}$ added to hyperedge $e_j \in \hat{\mathcal{E}}$. The attack is articulated as follows:  
\begin{align}\label{eq:attack_obj}
\min~{}\mathcal{L}_{\text{atk}}(f_{\theta^{*}}(\hat{\mathcal{G}})), \quad \text{s.t.}~\|\hat{\mathcal{G}}- \mathcal{G}\| \leq \Phi,
\end{align}
where $\theta^* = \underset{\theta}{\arg\min}\ \mathcal{L}_{atk}(f_{\theta}(\hat{\mathcal{G}}))$, represents the HGNNs trained on attacked data. $\Phi$ is the budget for the number of injected nodes, determined by the perturbation rate $\eta$ and the number of nodes $N$.
\subsubsection{Existing Challenges of Attack in HGNNs}
Existing attack methods for HGNNs focus on identifying target hyperedges for alteration or injecting nodes. Hypergraph modification attacks leverage model's gradient information to obtain target hyperedges, while hypergraph injection attacks rely on hypergraph structural characteristics. However, these methods rely on the specific information mechanisms of target HGNNs, overlooking the common vulnerability arising from the significant differences in hyperedge pivotality along aggregation paths in most HGNNs. This oversight limits the transferability and effectiveness of attacks.

Therefore, this study faces two key challenges: (1) investigating varying pivotality of hyperedges in information aggregation paths about HGNNs and developing an effective pivotal hyperedge identification mechanism to maximize the disruptive impact of injected malicious nodes; (2) leveraging the common vulnerabilities in the aggregation structures of diverse HGNNs to design a feature inverter, enabling precise disruption of aggregation paths while preserving the stealthiness of injected nodes.
\section{Method}
\begin{figure*}[!htbp]
  \centering
  \includegraphics[width=0.95\linewidth]{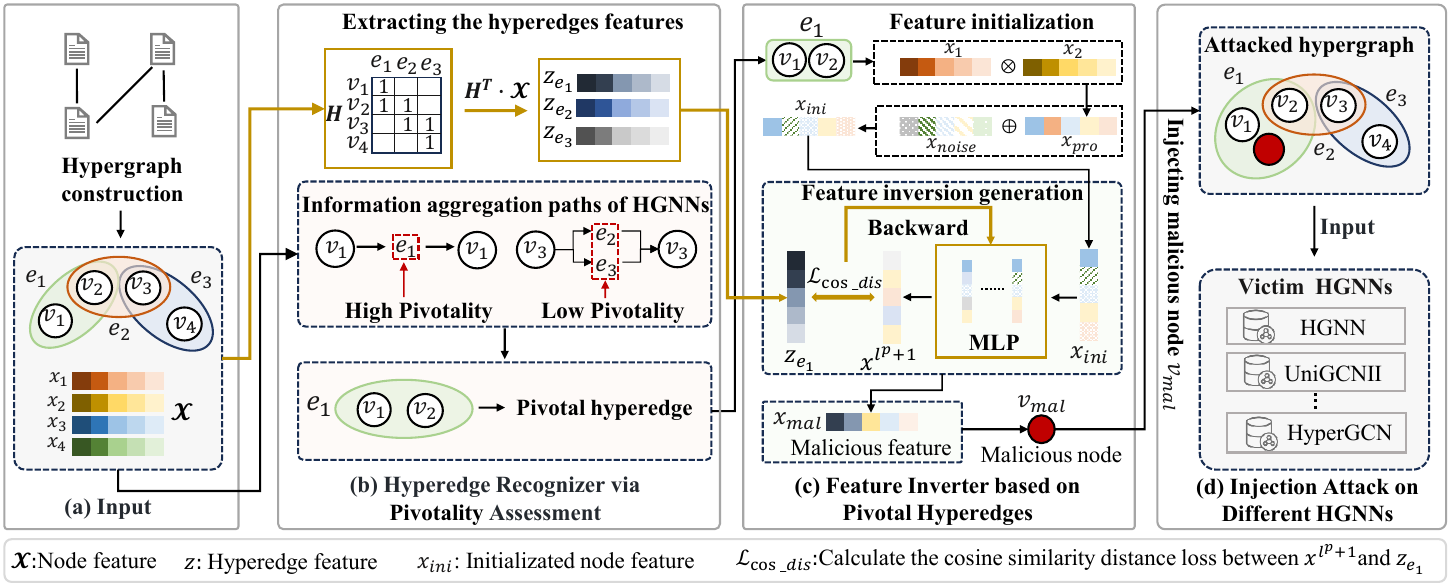}
  \caption{Framework of TH-Attack. (a) A hypergraph constructed from a regular graph as input for the model. (b) We propose a hyperedge recognizer via pivotality assessment to identify pivotal hyperedges. (c) Generating malicious nodes using a feature inverter based on pivotal hyperedges. (d) By injecting malicious nodes into the pivotal hyperedge to obtain an attacked hypergraph, which serves as input for different HGNNs and disrupts the performance of HGNNs.}
  \label{frame}
\end{figure*}
\subsubsection{Hyperedge Recognizer via Pivotality Assessment}
\begin{figure}[!htbp]
  \centering 
  \includegraphics[width=0.95\linewidth]{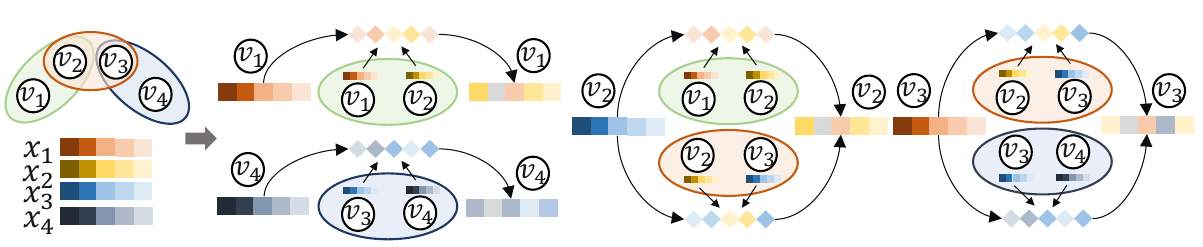}
  \caption{The information aggregation paths of HGNNs.}
  \label{aggregation}
\end{figure}

To tackle the first challenge of attacking HGNNs highlighted in this paper, we first analyze the feature aggregation process of HGNNs, as illustrated in Eq. (\ref{HGNN_eq}). Specifically, the node-hyperedge aggregation process is:
\begin{align}\label{node_hyperedge}
\mathbf{Z}^{(l)} = \sigma\left( D_{\mathcal{E}}^{-1} H^{\mathsf{T}} D_{\mathcal{V}}^{-\frac{1}{2}} \mathcal{X}^{(l)} \Theta^{(l)} \right),
\end{align}
here, $\mathbf{Z}^{(l)} \in \mathbb{R}^{|\mathcal{E}| \times |\mathcal{F}|}$ represents the hyperedge feature matrix at the $l$-th layer. This process is defined as $
\mathbf{z}_j^{(l)} = \frac{1}{|e_j|} \sum_{v_i \in e_j} \frac{1}{\sqrt{d_{v_i}}} \mathbf{x}_i^{(l)} \Theta^{(l)}.
$
Furthermore, the hyperedge-node aggregation process is:  
\begin{align}
\mathcal{X}^{(l+1)} = \sigma\left( D_{\mathcal{V}}^{-\frac{1}{2}} H \mathcal{W} \mathbf{Z}^{(l)} \right).
\end{align} 
In component form, this is expressed as:  
$
\mathbf{x}_i^{(l+1)} = \frac{1}{\sqrt{d_{v_i}}} \sum_{e_j \ni v_i} w_{e_j} \mathbf{z}_j^{(l)}.
$
From the aforementioned feature aggregation processes, it is evident that the update of node feature representations necessitates traversing the node-hyperedge-node aggregation pathway. However, within hypergraph structures, nodes are not limited to belonging to just one hyperedge. Some nodes may acquire higher-order feature representations through multiple hyperedges, while others may only do so through a few. As illustrated in Figure \ref{aggregation}, nodes $v_{1}$ and $v_{4}$ obtain higher-order features solely via hyperedges $e_{1}$ and $e_{2}$, respectively. In contrast, nodes $v_{2}$ and $v_{3}$ acquire such features through two hyperedges.  

Consequently, these hyperedges within aggregation pathways involving a limited number of hyperedges have higher pivotality. Disrupting these hyperedge could impair the information propagation process, leading to error results. On this basis, we propose a hyperedge recognizer via pivotality assessment. For a node $v_{i} \in \mathcal{V}$, we define its isolation degree as its hyperdegree:
\begin{align}
d_h(v_i) = |\{ e_j \in \mathcal{E} \mid v_i \in e_j \}|.
\end{align}
Here, node $v_{j}$ is in pivotal hyperedges if its isolation degree is in pivotality level $\tau$: $d_h(v_i) \leq \tau$. We furthermore analyze the impact of different $\tau$ values on attack performance in the experimental section.

The theoretical foundation of the aforementioned pivotality assessment lies in the perturbation propagation characteristics of aggregation paths. And we reveal the following fundamental principles.

\begin{theorem} (Perturbation Amplification in High Pivotality Hyperedges)
When node $v_i$ aggregates information through the highly pivotal hyperedges (i.e., $d_h(v_i) \leq \tau$), the lower bound of its feature perturbation is given by:
\begin{align}
\|\Delta \mathbf{x}_i^{(l+1)}\|_2 \geq \frac{1}{\sqrt{d_{v_i}}} \min{e_j \ni v_i} w_{e_j} \cdot \|\Delta \mathbf{z}_j^{(l)}\|_2.
\end{align}
\end{theorem}

\textit{Proof:} Considering the most vulnerable scenario where information propagates through higher pivotality hyperedges, the perturbation amplification effect follows from:
\begin{align}
\| \sum_{e_j \ni v_i} \frac{w_{e_j}}{\sqrt{d_{v_i}}} \Delta \mathbf{z}_j^{(l)} \|_2 \nonumber
&\geq \frac{1}{\sqrt{d_{v_i}}} \min_{e_j \ni v_i} w_{e_j} \cdot \|\Delta \mathbf{z}_j^{(l)}\|_2,
\end{align}
where $\|\Delta \mathbf{x}_i^{(l+1)}\|_2=\| \sum_{e_j \ni v_i} \frac{w_{e_j}}{\sqrt{d_{v_i}}} \Delta \mathbf{z}_j^{(l)} \|_2$.
\begin{theorem} (Perturbation Amplification in Low Pivotality Hyperedges)
When node $v_k$ aggregates information through a lower pivotality hyperedges (i.e., $d_h(v_k) > \tau$), the upper bound of its feature perturbation is given by:
\begin{align}
\|\Delta \mathbf{x}_k^{(l+1)}\|_2 \leq \frac{1}{\sqrt{d_{v_k}}} \sum_{e_j \ni v_k} w_{e_j} \|\Delta \mathbf{z}_j^{(l)}\|_2.
\end{align}
\end{theorem}

\textit{Proof:} Applying the triangle inequality to the aggregation formula:
\begin{align}
\| \sum_{e_j \ni v_k} \frac{w_{z_j}}{\sqrt{d_{v_k}}} \Delta \mathbf{z}_j^{(l)} \|_2 \leq \frac{1}{\sqrt{d_{v_k}}} \sum_{e_j \ni v_k} w_{e_j} \|\Delta \mathbf{z}_j^{(l)}\|_2.
\end{align}

Theorems 1 and 2 collectively establish the theoretical basis for the pivotality assessment mechanism. Nodes with higher pivotality hyperedges in aggregation paths exhibit perturbation amplification, where attacking associated hyperedges amplifies feature perturbation $\|\Delta \mathbf{x}_i\|_2$ by at least $\frac{1}{\sqrt{d_{v_i}}} \min w_{e_j}$. Conversely, nodes in hyperedges with lower pivotality demonstrate perturbation attenuation, where disturbance energy is dissipated through redundant pathways. This validates the scientific rationale for selecting pivotal nodes via the pivotality level $\tau$ and constructing the attack pivotal hyperedges $\mathcal{E}_{\text{all\_piv}}$ from their associated hyperedges.

Therefore, we obtain the pivotal hyperedges in information aggregation paths of HGNNs:
\begin{align}\label{target}
\mathcal{E}_{\text{all\_piv}} = \{ e_j \in \mathcal{E} \mid \exists v_i \in e_j \text{ such that } d_h(v_i) \leq \tau \}.
\end{align}
Then, we randomly select a corresponding number of hyperedges as final pivotal hyperedges \(\mathcal{E}_{\text{A-P}}\) from \(\mathcal{E}_{\text{all\_piv}}\) based on this budget, typically not exceeding 5\% of all nodes.
\subsubsection{Feature Inverter based on Pivotal Hyperedges}
In the information aggregation path of HGNNs, hyperedges act as critical intermediaries within the ``node-hyperedge-node" feature propagation chain, playing a central role in regulating the flow of information. Traditional feature generators focus on learning the feature distribution of target nodes to generate features that conform to this distribution, while neglecting the propagation characteristics of the target nodes. To effectively disrupt the information propagation properties of pivotal hyperedges, we propose a feature inverter based on pivotal hyperedges.

For each pivotal hyperedge $e_j \in \mathcal{E}_{\text{A-P}}$, we first generate a initial feature $\mathbf{x}_{ini}^{(j)}$ by combining the internal node features of the hyperedge with random noise:
\begin{align}
\mathbf{x}_{ini}^{(j)} = \mathbf{x}_{pro}^{(j)} \oplus \mathcal{N}(0,\mu^2),
\end{align}
where $\mathbf{x}_{pro}^{(j)} = \prod_{v_i \in e_j} \mathbf{x}_i$ represents an element-wise product along the feature dimension, exemplified by $e_{1}=\{v_{1},v_{2}\}$ and $\mathbf{x}_{pro}^{(1)}=\mathbf{x}_{1}\otimes \mathbf{x}_{2}$ in Figure \ref{frame}(c). $\mathcal{N}(0, \mu^2)$ represents Gaussian noise with zero mean and variance $\mu^2$. This design enhances feature diversity while preserving statistical correlations with original hyperedge features. The $\mathbf{x}_{ini}^{(j)}$ matches the dimensionality of node features, providing an initial condition for subsequent feature generation.

Furthermore, we utilze the Multi-Layer Perceptron (MLP) \cite{riedmiller2014multi} to enhance the initial confusion features, maximizing the feature confusion effect while maintaining the concealment of the topological structure. The formula is defined as follows:
\begin{align}
\mathbf{x}^{(l^{p}+1)}= \sigma(\mathbf{W}^{(l^{p})} \mathbf{x}^{(l^{p})} + \mathbf{b}).
\end{align}
Herein, $\mathbf{W}^{(l^{p})}$ is the weight of the $l^{p}$-th layer. When $l^{p}=0$, $\mathbf{x}^{(l^{p})}$ is the $\mathbf{x}_{ini}^{(j)}$. $\mathbf{b}$ is the offset and $\sigma$ is the LeakyReLU activation function. Subsequently, the generated malicious feature $\mathbf{x}_{mal}^{(j)}=Softmax(\mathbf{x}^{(l^{p})})$ is obtained.  

To maximize the semantic divergence between the generated features and the pivotal hyperedge, while constraining the feature deviation using a differentiable threshold regularization term, we design a loss function:
\begin{align}
\mathcal{L}_{cos\_dis} = \cos(\mathbf{x}_{mal}^{(j)}, \mathbf{z}_{e_{j}}) + \lambda \cdot \mathcal{L}_{reg},
\end{align}
where $\mathcal{L}_{reg} = \max(  \cos(\mathbf{x}_{mal}^{(j)},\mathbf{z}_{e_{j}}) - t, 0 )$, and \(t\) is the similarity threshold. $\mathbf{z}_{e_{j}} \in \mathbf{Z}$ is the hyperedge feature obtained by $\mathbf{Z}=H^{\mathsf{T}} \cdot \mathcal{X}$, which is a simplified version derived from Eq. (\ref{node_hyperedge}) relies on the hypergraph structure, without knowledge of model parameters. The similarity constraint ensures that the generated features effectively disrupt the pivotal hyperedge, while the $\lambda \cdot \mathcal{L}_{reg}$ promotes feature smoothness to maintain stealthiness. Model parameters of inverter are updated via backpropagation, allowing the generated features to evolve along the direction of the loss gradient. The hyperparameter $\lambda$ controls the trade-off between attack strength and stealthiness. Consequently, we obtain the malicious nodes $v_{mal}^{(j)}$ with malicious feature $\mathbf{x}_{mal}^{(j)}$. 

\subsubsection{Injection Attack on Different HGNNs}
For each pivotal hyperedge $e_{j}\in \mathcal{E}_{\text{A-P}}$, we generate a malicious node $v_{mal}^{(j)}$, which is then injected into the pivotal hyperedge $e_{j}$, as illustrated in Figure \ref{frame}(d). Consequently, for the hypergraph $\mathcal{G} = (\mathcal{V}, \mathcal{E})$, all selected pivotal hyperedges are denoted as $\mathcal{E}_{\text{A-P}}$ according to Eq. (\ref{target}). Thus, the attacked hypergraph $\hat{\mathcal{G}} = (\hat{\mathcal{V}}, \hat{\mathcal{E}})$ is obtained by injecting malicious nodes $\mathcal{V}_{mal}$ generated through feature inverter into the selected pivotal hyperedges $\mathcal{E}_{\text{A-P}}$, where $\hat{\mathcal{V}} = \mathcal{V} \cup \mathcal{V}_{mal}$, and $\hat{\mathcal{E}} = \{ e_{j} \cup v_{mal}^{(j)} \mid e_{j} \in \mathcal{E} \}$, denoting nodes injected into hyperedge $e_{j}=\{v_{1},v_{2}\}\rightarrow\{v_{1},v_{2},v_{mal}^{(j)}\}$. The incidence matrix $\hat{H} \in \mathbb{R}^{(|\hat{\mathcal{V}}|+m) \times |\hat{\mathcal{E}|}}$ is expanded as:
\begin{align}
\quad \hat{H}_{ij} = \left\{
\begin{aligned}
&1, ~~~~ \text{if } v_i \in e_j \text{ and }  v_i \in \mathcal{V}, \\
&1, ~~~~ \text{if } v_i \in \mathcal{V}_{mal} \text{ and } e_j \in \mathcal{E}_{\text{A-P}}, \\
&0, ~~~~ \text{otherwise}.
\end{aligned}
\right.
\end{align}

It is evident that the node dimension of $\widehat{H}$ increases by $m$, while the hyperedge dimension remains unchanged. Additionally, the feature matrix $\mathcal{X}$ has been augmented with the injected malicious nodes features $\mathcal{X}_{mal}$ for malicious nodes $\mathcal{V}_{mal}$. 
Therefore, we obtain the attacked data for the attack on HGNNs, which includes the attacked incidence matrix $\widehat{H}$ and the disturbed node features $\widehat{\mathcal{X}} = \{\mathcal{X}, \mathcal{X}_{mal}\}$. 

It is crucial to emphasize that our injection attack operates under a black-box scenario, achieving transferable attacks by feeding the attacked hypergraph into diverse HGNNs. The efficacy of this black-box attack stems from precise exploitation of inherent vulnerabilities in HGNNs. Attackers require no knowledge of the target model's specific parameters $\theta$ or architectural details, leveraging the attacked hypergraph $\hat{\mathcal{G}}$ as the input of HGNNs. When $\hat{\mathcal{G}}$ is processed by any HGNN model $f_{\theta}$, its aggregation mechanism:
\begin{align}  
\hat{\mathcal{X}}^{(l+1)} = \sigma\Big( \hat{D}_{\mathcal{V}}^{-\frac{1}{2}} \hat{H} \mathcal{W} \hat{D}_{\mathcal{E}}^{-1} \hat{H}^{\mathsf{T}} \hat{D}_{\mathcal{V}}^{-\frac{1}{2}} \hat{\mathcal{X}}^{(l)} \Theta^{(l)} \Big).  
\end{align}  
The spectral radius of the hyperedge degree matrix satisfies $\rho(\hat{D}_{\mathcal{E}}^{-1}) \leq \rho(D_{\mathcal{E}}^{-1}) + \mathcal{O}\left(\frac{M}{\min_j [D_{\mathcal{E}}]_{jj}^2}\right)$, leading to miscalibration in information weight allocation. The gradient propagation of malicious features is constrained by the spectral norm of the operator $\hat{D}_{\mathcal{V}}^{-1/2} \hat{H} \mathcal{W} \hat{D}_{\mathcal{E}}^{-1}$: $\|\nabla_{\mathbf{x}_{mal}} \hat{\mathcal{X}}^{(l)}\|_F \leq C \cdot \|\hat{D}_{\mathcal{V}}^{-1/2} \hat{H} \mathcal{W} \hat{D}_{\mathcal{E}}^{-1}\|_2 \cdot \|\hat{\mathcal{X}}^{(l-1)}\Theta^{(l-1)}\|_F$. Such attacks consistently induce performance degradation across various models (HyperGCN, HGNN,$\ldots$, UniGCNII), indicating that their generalization stems from perturbations to the hypergraph structure rather than specific model parameters. 
\section{Experiments}
\begin{table*}[t]\small
\centering
\setlength{\tabcolsep}{5.7pt}
    \scalebox{0.9}{\begin{tabular}{llcccccccccccccccc}
        \toprule
        \multirow{2}{*}{Datasets}&\multirow{2}{*}{Models}&\multicolumn{2}{c}{Clean}&\multicolumn{2}{c}{Random} &\multicolumn{2}{c}{DICE} &\multicolumn{2}{c}{FGA} &\multicolumn{2}{c}{IGA} & \multicolumn{2}{c}{IE-Attack}&\multicolumn{2}{c}{TH-Attack}\\
        
        \cmidrule(lr){3-4} \cmidrule(lr){5-6}\cmidrule(lr){7-8}\cmidrule(lr){9-10}\cmidrule(lr){11-12}\cmidrule(lr){13-14}\cmidrule(lr){15-16}
&&$Acc$&$F1$&$Acc$&$F1$&$Acc$&$F1$&$Acc$&$F1$&$Acc$&$F1$&$Acc$&$F1$&$Acc$&$F1$\\
        \hline
        \multirow{5}{*}{Cora}& CEGCN&74.84&72.65&72.06&69.43&71.95&68.99&72.00&69.44&72.26&69.88&\underline{70.73}&\underline{67.92}&\textbf{67.27}&\textbf{63.71}\\
        & CEGAT&74.80&72.44&71.99&68.85&72.73&70.21&72.99&70.67&71.68&69.65&\underline{71.44}&\underline{68.45}&\textbf{66.98}&\textbf{60.83}\\
        & HGNN&76.41&73.97&74.71&72.19&74.45&71.66&74.41&72.19&73.84&71.33&\underline{73.20}&\underline{69.33}&\textbf{36.08}&\textbf{22.94}\\
        & HyperGCN&75.95&70.98&73.14&69.35&73.93&69.81&72.62&68.55&71.09&66.16&\underline{68.37}&\underline{59.86}&\textbf{31.55}&\textbf{16.09}\\
        & UniGCNII&80.08&77.99&\underline{75.82}&\underline{72.84}&77.23&74.61&76.65&74.67&76.01&73.60&76.57&74.15&\textbf{39.42}&\textbf{21.78}\\
        \hline
        \multirow{5}{*}{Cora-CA}& CEGCN&75.92&73.64&72.62&70.76&\underline{72.38}&\underline{69.69}&74.28&72.43&74.71&72.59&74.80&73.25&\textbf{50.39}&\textbf{36.71}\\
        & CEGAT&76.16&74.69&\underline{73.79}&\underline{71.10}&73.77&71.57&73.98&72.33&74.57&72.51&76.01&73.50&\textbf{60.05}&\textbf{51.31}\\
        & HGNN&82.45&80.34&78.90&76.91&79.08&76.98&78.55&76.74&\underline{77.13}&\underline{74.94}&80.79&78.72&\textbf{32.03}&\textbf{18.59}\\
        & HyperGCN&76.32&71.99&75.61&72.36&75.75&71.98&72.45&70.23&\underline{68.93}&\underline{65.08}&73.82&65.95&\textbf{17.72}&\textbf{13.02}\\
        & UniGCNII&84.68&83.18&79.53&78.21&80.15&78.14&80.57&79.17&\underline{79.07}&\underline{77.49}&83.95&82.43&\textbf{32.72}&\textbf{13.42}\\
        \hline
        \multirow{5}{*}{Citeseer}& CEGCN&68.79&64.25&\underline{66.12}&\underline{61.66}&66.20&62.11&67.62&62.69&66.19&61.87&68.68&63.37&\textbf{52.21}&\textbf{44.03}\\
        & CEGAT&70.48&65.15&\underline{66.13}&\underline{61.57}&67.60&62.68&67.05&62.68&66.26&61.05&68.77&63.59&\textbf{62.85}&\textbf{56.63}\\
        & HGNN&71.01&66.45&67.92&63.42&67.80&62.91&67.28&\underline{62.61}&\underline{67.01}&62.77&68.03&63.77&\textbf{24.17}&\textbf{17.94}\\
        & HyperGCN&70.78&66.51&68.25&63.85&67.96&64.32&67.83&\underline{63.01}&\underline{67.75}&63.59&68.75&64.78&\textbf{20.59}&\textbf{12.38}\\
        & UniGCNII&72.18&67.62&69.85&65.36&69.89&65.54&71.44&66.82&\underline{69.23}&\underline{64.80}&70.95&65.88&\textbf{27.00}&\textbf{15.85}\\
        \hline
        \multirow{5}{*}{Pubmed}& CEGCN&86.13&85.72&\underline{83.27}&\underline{82.73}&83.36&82.82&83.29&82.75&83.39&83.08&85.72&85.34&\textbf{48.87}&\textbf{36.81}\\
        & CEGAT&85.81&85.45&83.24&82.81&83.29&82.78&83.09&\underline{82.46}&\underline{82.25}&81.86&85.89&85.60&\textbf{52.34}&\textbf{39.58}\\
        & HGNN&84.28&84.16&\underline{80.99}&\underline{80.73}&81.29&80.92&81.99&81.68&81.60&81.42&84.53&84.48&\textbf{40.96}&\textbf{30.51}\\
        & HyperGCN&76.33&73.29&73.28&69.02&\underline{70.29}&\underline{65.83}&83.23&82.99&74.17&70.82&75.69&72.39&\textbf{35.14}&\textbf{28.00}\\
        & UniGCNII&87.86&87.64&84.22&83.81&\underline{83.95}&\underline{83.46}&85.09&84.70&84.15&83.75&87.62&87.40&\textbf{44.13}&\textbf{31.96}\\
        \hline
        \multirow{5}{*}{DBLP}& CEGCN&87.59&86.98&84.14&83.57&84.15&83.79&84.05&\underline{83.55}&\underline{84.03}&83.72&87.60&87.05&\textbf{74.22}&\textbf{70.10}\\
        & CEGAT&88.37&87.85&84.64&83.86&84.77&83.89&84.05&83.55&\underline{83.74}&\underline{83.25}&88.01&87.50&\textbf{77.62}&\textbf{75.64}\\
        & HGNN&91.03&84.15&86.54&82.37&83.78&81.62&83.19&80.67&82.73&\underline{80.13}&\underline{81.37}&84.63&\textbf{64.97}&\textbf{59.46}\\
        & HyperGCN&89.54&89.10&83.84&82.71&\underline{81.72}&\underline{78.79}&85.85&85.35&81.83&80.26&82.64&79.37&\textbf{46.23}&\textbf{14.82}\\
        & UniGCNII&91.89&91.61&87.89&87.22&87.79&86.98&88.40&87.97&\underline{87.52}&\underline{86.66}&88.31&89.16&\textbf{38.76}&\textbf{19.87}\\
        \hline
        \multirow{5}{*}{ModelNet40}& CEGCN&89.82&86.99&83.92&78.61&83.74&78.59&83.61&78.32&\underline{82.39}&\underline{75.94}&85.94&81.13&\textbf{78.24}&\textbf{67.49}\\
        & CEGAT&92.31&89.78&87.85&82.23&88.16&82.52&88.20&84.10&\underline{84.07}&\underline{78.00}&88.34&83.99&\textbf{78.18}&\textbf{66.09}\\
        & HGNN&95.48&93.93&89.13&83.04&89.24&\underline{82.98}&89.73&\underline{84.67}&89.12&84.97&93.82&91.72&\textbf{72.91}&\textbf{62.60}\\
        & HyperGCN&86.45&81.48&83.88&71.67&83.67&71.35&\underline{77.36}&\underline{63.05}&80.76&66.31&80.53&67.71&\textbf{50.11}&\textbf{24.16}\\
        & UniGCNII&97.86&97.03&93.45&\underline{88.89}&93.51&89.44&\underline{93.36}&90.71&93.48&89.38&96.65&95.67&\textbf{53.50}&\textbf{30.14}\\
        \bottomrule
    \end{tabular}}
    \caption{Comparison of \textit{Accuracy} (\textit{Acc}) (\%) and \textit{Macro\_F1} (\textit{F1}) (\%) of TH-Attack and baselines. The results are averaged over 10 runs on different random seeds. The best values are in \textbf{bold}, and the second-best values are \underline{underlined}.}
    \label{performance_all}
\end{table*}

\subsection{Experimental Setting}
\subsubsection{Datasets} Six benchmark datasets are adopted in this paper, including citation networks (Cora, Citeseer, and Pubmed) \cite{yadati2019hypergcn}, Co-authorship networks (Cora-CA, DBLP) \cite{chien2021you}. Additionally, we evaluate our approach on the ModelNet40 \cite{wu20153d}, which is widely used in computer vision and graphics. The hypergraph construction process follows the guidelines provided in HGNNs \cite{feng2019hypergraph,yang2022semi}. In addition, the partitioning strategy of the dataset follows the Allset \cite{chien2021you}, which are divided into training/validation/test sets. 

\subsubsection{Parameter Setting} 
This study examines the impact of four key parameters on the experimental outcomes: the attack budget $\eta$, the pivotality level $\tau$, and the hyperparameters $\lambda$ and $t$ used to constrain the magnitude of feature deviation. A detailed sensitivity analysis of these parameters is provided in the experimental section. Additionally, various parameters used in testing attacks on different HGNNs refer to the Allset \cite{chien2021you}.

\subsubsection{Evaluating Metrics} This paper utilizes the \textit{Accuracy} and \textit{Macro\_F1} as evaluation metrics for the performance of TH-Attack. The \textit{Accuracy} and \textit{Macro\_F1} indicates the node classification performance of HGNNs, where a lower \textit{Accuracy} and \textit{Macro\_F1} signifies a more effective attack.

\subsubsection{Baselines} We set up five baselines, including random methods (Random, DICE \cite{Waniek:dice2018}), gradient methods (FGA \cite{chen2018fast}, IGA \cite{wu2019adversarial}), and hypergraph injection methods (IE-Attack \cite{he2025hypergraph}. Additionally, both the TH-Attack and baselines are conducted in a black-box attack scenario against HGNNs (CEGCN and CEGAT \cite{chien2021you}, HGNN \cite{feng2019hypergraph}, HyperGCN \cite{yadati2019hypergcn}, UniGCNII \cite{huang2021unignn}). All experiments are conducted on a workstation equipped with four NVIDIA RTX 3090 GPUs, which are conducted under the same parameters settings. 
\subsection{Model Performance and Parameter Analysis}
\subsubsection{Performance Comparison with State-of-the-art Methods}
Table \ref{performance_all} presents the node classification results of the proposed TH-Attack compared to several baselines across six datasets targeting five HGNNs. The Random and DICE methods employ a strategy of randomly selecting hyperedges for node injection, whereas FGA and IGA utilize gradient guidance to select hyperedges for node injection. All baselines are under a black-box poisoning scenario similar to proposed TH-Attack. It highlights that the original design of IE-Attack was for gray-box evasion attacks. For the sake of comparability, we adapt it to a black-box poisoning scenario, aligning it with TH-Attack. 

From Table \ref{performance_all}, TH-Attack demonstrates significantly superior attack effectiveness across multiple datasets and models compared to baseline methods. For instance, on the Cora, TH-Attack reduces the $Accuracy$ of HGNN from 76.41\% to 36.08\%, a decrease of 40.33 percentage points, which far exceeds the reduction achieved by random methods (Random/DICE) of approximately 2-3 percentage points and gradient-guided methods (FGA/IGA) of about 1-2 percentage points. Notably, on complex datasets such as ModelNet40, baselines induce only limited performance degradation. Furthermore, TH-Attack exhibits pronounced advantages in cross-model transferability. For instance, IE-Attack mainly focuses on HyperGCN, demonstrating slightly higher effectiveness on the Cora dataset for HyperGCN compared to other HGNNs. Consequently, TH-Attack's strategy of injecting nodes generated by the feature inverter into pivotal hyperedges proves to be highly effective across various HGNNs, showing exceptional attack performance and high transferability.

\subsubsection{Analysis of the Perturbation Rate $\eta$}
\begin{figure}[!htbp]
  \centering
  \includegraphics[width=0.90\linewidth]{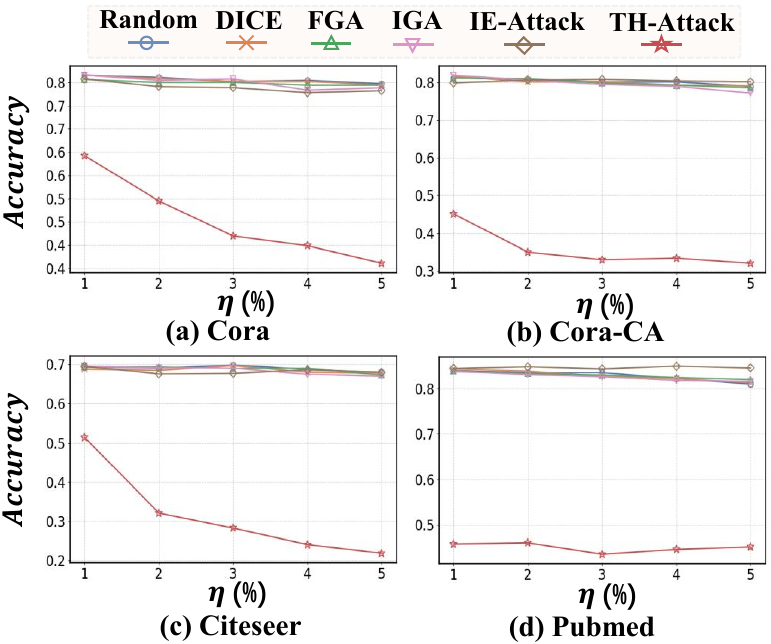}
  \caption{The \textit{Accuracy} of TH-Attack compared to baselines under different $\eta$ against HGNN.}
  \label{ptb_rate}
\end{figure}

We examine results for $\eta$ ranging from 1\% to 5\%, as illustrated in Figure \ref{ptb_rate}. Overall, with $\eta$ increases, the performance of HGNN under TH-Attack deteriorates significantly more than with baselines. Notably, this experiment demonstrates that our TH-Attack maintains superior attack performance even with minimal node injections. For instance, when $\eta=1\%$, only 23 nodes are injected into the Cora, resulting in the 0.1717 drop in HGNN $Accuracy$ attacked by TH-Attack, compared to a maximum 5\% reduction with baselines. Therefore, our strategy of injecting carefully crafted malicious nodes into pivotal hyperedges along information aggregation paths significantly enhances hypergraph attack performance.
\subsubsection{Analysis of the Pivotality Level $\tau$}
We evaluate attack performance across different pivotality level $\tau$, as shown in Table \ref{performance_vulnerability}. The level $\tau$ is determined by the number of hyperedges the node traverses in the aggregation paths. For example, if the hyperedge count ranges from $\{1, 2, \ldots, 12\}$, it is divided into six segments. At level $\tau=1$, the hyperedge count is $\{1, 2\}$, indicating the highest pivotality. Higher $\tau$ correspond to lower pivotality. The results indicate that when $\tau \geq 2$, the $Accuracy$ of HGNN generally increases across datasets, suggesting a decline in attack performance. Nevertheless, this experiment confirms that injecting nodes into pivotal hyperedges within information aggregation paths enhances the attack's performance against HGNN. This is because smaller $\tau$ values indicate higher pivotality hyperedges in information aggregation path, resulting in more malicious interference with the aggregation process.
\begin{table}[!htbp]\small
\centering
\setlength{\tabcolsep}{4pt}
    \begin{tabular}{lcccccc}
        \toprule
        \multirow{1}{*}{Datasets}&\multicolumn{1}{c}{$\tau = 1$}&\multicolumn{1}{c}{$\tau = 2$} &\multicolumn{1}{c}{$\tau = 3$} &\multicolumn{1}{c}{$\tau = 4$} &\multicolumn{1}{c}{$\tau = 5$} & \multicolumn{1}{c}{$\tau = 6$}\\
        \hline
        \multirow{1}{*}{Cora}
        &39.15&\textbf{37.16}&39.90&43.63&39.95&42.13\\
        \multirow{1}{*}{Cora-CA}
        &30.00&\textbf{29.95}&31.15&32.36&30.54&30.78\\
        \multirow{1}{*}{Citeseer}
        &22.52&\textbf{21.59}&22.12&24.65&25.16&24.19\\
        \multirow{1}{*}{Pubmed}&41.10&\textbf{40.45}&42.47&42.98&43.16&43.97\\
        \bottomrule
    \end{tabular}
    \caption{The \textit{Accuracy} (\%) of TH-Attack under different pivotality level $\tau$ against HGNN.}
    \label{performance_vulnerability}
\end{table}
\subsubsection{Analysis of the Hyperparameters $\lambda$ and $t$}
\begin{figure}[t]
  \centering
  \includegraphics[width=0.90\linewidth]{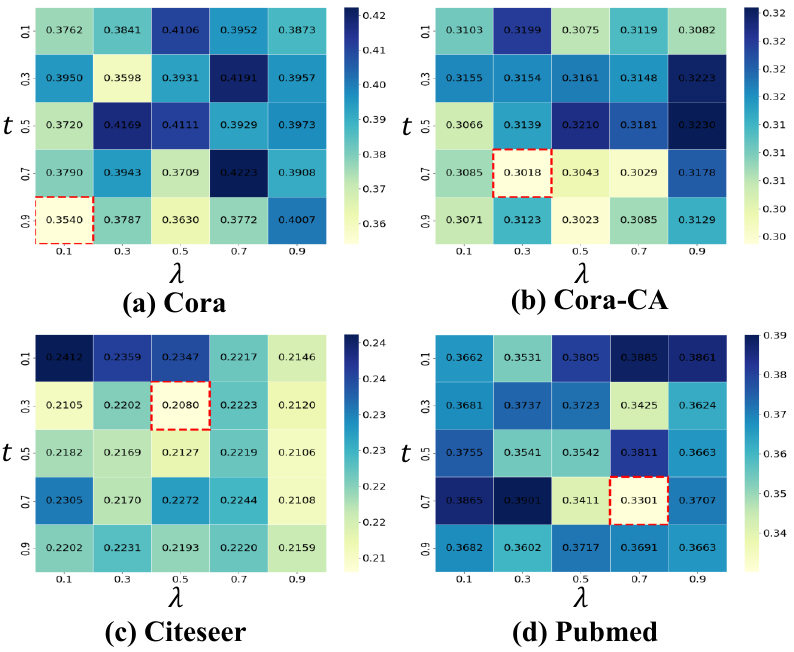}
  \caption{The \textit{Accuracy} of TH-Attack under different $\lambda$ and $t$ against HGNN.}
  \label{lamda_t}
\end{figure}

We conduct attack performance concerning the regularization coefficient $\lambda$ and similarity threshold $t$, as shown in Figure \ref{lamda_t}. TH-Attack shows optimal attack performance in $\lambda=0.1$ and $t=0.9$ (red box), achieving a minimum \textit{Accuracy} of 0.3540 on Cora. This optimal combination uses a low $\lambda$ to enhance attack intensity by reducing regularization constraints, and a high $t$ to preserve crucial feature perturbations by relaxing similarity thresholds. Therefore, $\lambda$ and $t$ maximize semantic differences while preventing excessive suppression, achieving a balance between attack effectiveness and stealth.

\subsection{Ablation Study and Analysis}
We conduct ablation experiments by removing the ``Hyperedge Recognizer (HR)", ``Feature Inverter (FI)", and ``Cosine Distance Loss (CDL)". Table \ref{performance_ablation} indicates that excluding these three strategies leads to a decline in the TH-Attack's performance, resulting in higher $Accuracy$. Notably, the absence of ``CDL" has a significant impact on the TH-Attack's performance. TH-Attack enhances its attack performance and improves cross-model transferability by maximizing the influence of malicious nodes generated by the inverter on pivotal hyperedges along information aggregation paths. Therefore, the ablation experiments confirm the remarkable effectiveness and transferability of the proposed TH-Attack across different HGNNs.
\begin{table}[htbp]\small
  \centering
\setlength{\tabcolsep}{3pt}
  \begin{threeparttable}
  
    \begin{tabular}{llccccc}
    \toprule
    \multicolumn{1}{c}{Modules}
    &Models&\multicolumn{1}{c}{Cora}&\multicolumn{1}{c}{Cora-CA}&\multicolumn{1}{c}{Citeseer}&\multicolumn{1}{c}{Pubmed}\cr
    \hline
    \multirow{3}{*}{w/o HR}
    
    &HGNN &41.19&38.30&28.69&45.16\cr
    &HyperGCN &34.87&24.51&21.43&36.47\cr
    &UniGCNII &43.48&40.00&34.05&47.14\cr
    \midrule
    \multirow{3}{*}{w/o FI} 
    &HGNN &61.80&58.40&54.26&44.25\cr
    &HyperGCN &38.65&26.57&43.93&38.97\cr
    &UniGCNII &73.85&77.72&66.94&47.95\cr
    \midrule
    \multirow{3}{*}{w/o CDL}
    &HGNN &61.05&59.31&54.45&55.85\cr
    &HyperGCN &59.42&54.66&52.64&52.76\cr
    &UniGCNII &59.80&60.34&66.06&46.40\cr
    \midrule
    \multirow{3}{*}{TH-Attack}
    &HGNN &\textbf{36.08}&\textbf{32.03}&\textbf{24.17}&\textbf{40.96}\cr
    &HyperGCN &\textbf{31.55}&\textbf{17.02}&\textbf{20.59}&\textbf{35.14}\cr
    &UniGCNII &\textbf{39.42}&\textbf{32.72}&\textbf{27.00}&\textbf{44.13}\cr
    \bottomrule
    \end{tabular}
    \caption{The \textit{Accuracy} (\%) of variants for TH-Attack on different HGNNs. w/o means remove this strategy.}
    \label{performance_ablation}
    \end{threeparttable}
\end{table}

\section{Conclusion}
This paper explores a common vulnerability in HGNNs due to the varying pivotality of hyperedges in the information aggregation path and introduces the TH-Attack. TH-Attack uses a hyperedge recognizer to identify pivotal hyperedges and a feature inverter to generate malicious nodes, which are then injected into these hyperedges to enhance attack transferability and effectiveness. Experiments on six real-world datasets across five HGNNs show that TH-Attack outperforms baselines. Future work will focus on attack strategies for dynamic HGNNs and cross-level knowledge techniques.

\section{Acknowledgments}
This work was supported in part by the National Natural Science Foundation of China (62572400,62472117), the Guangdong Basic and Applied Basic Research Foundation (2025A1515010157), the Science and Technology Projects in Guangzhou (2025A03J0137), the National Natural Science Foundation of China, Regional Science Foundation Project, 62466042.
\bibliography{aaai2026}
\end{document}